# Magnonic Hong-Ou-Mandel Effect

Mikhail Kostylev

*Department of Physics, the University of Western Australia*

*Crawley 6009 WA, Australia*

Abstract

We carried out numerical simulations of propagation of spin waves (magnons in quantum language) in a yttrium-iron garnet film. The numerical model is based on an original formalism. We demonstrated that a potential barrier for magnons, created by an Orsted field of a dc current flowing through a wire sitting on top of the film, is able to act as an electrically controlled partly transparent mirror for the magnons. We found that the mirror transparency can be set to 50% by properly adjusting the current strength, thus creating a semi-transparent mirror. A strongt Hong-Ou-Mandel Effect for single magnons is expected in this configuration. The effect must be seen as two single magnons, launched simultaneously into the film from two transducers located from the opposite sides of the mirror, creating a two-microwave-photon state at the output port of one of the transducers. The probability of seeing those two-photon states at the output port of either transducer must be the same for both transducers.

## I. Introduction

The Hong-Ou-Mandel Effect (HOME) represents a particular manifestation of the property of indistinguishable bosonic particles to bunch. It has been studied in detail for optical photons and is seen as a specific interference pattern, which pairs of photons generate while incident on a 50:50 lossless optical beam splitter [1]. If two indistinguishable optical photons are simultaneously incident on the beam splitter, each onto its own input port, both of them always exit the splitter together through the same output port, but never separately with one photon through each of the two output ports [2,3]. Later on it was theoretically shown that if the beam splitter is lossy the same interference pattern must remain for coincidence photons, i.e. when both photons survive the transmission through the lossy medium.

Very recently, the same effect was demonstrated in Quantum Acoustics for pairs of travelling phonons. Surface Acoustic Waves (SAW) in a LiNbO3 plates were employed, in order to observe this effect [4]. In the present work, we make a theoretical prediction that the same effect can be seen for travelling microwave magnons in single-crystal yttrium iron garnet (YIG) films. Central for observation of HOME is the availability of a semi-transparent mirror for particular (quasi)-particles, for which the effect is to be observed. In this work, we propose a partly transparent mirror for magnons, the degree of transparency of which can be controlled electrically, investigate its properties theoretically and consider its usefulness for observing HOME for travelling magnons.

Magnons (or magnon-polaritons) are bosonic quasi-particles that represent quanta of spin waves (SW). Quantum Magnonics has attracted a lot of attention recently, see Ref.[5] for a recent review on that matter and references therein. For instance, propagation of SW at millikelvin temperatures was successfully demonstrated [6,7] and shown that travelling single magnons can be employed to convert microwave photons to optical photons [8] and that parametrically squeezed states of travelling magnons can be created [9]. Classically, spin waves represent collective precession of electron spins in a magnetic material. The collective character of the spin dynamics is due to coupling of spins at nearby crystal lattice sites by exchange and dipole-dipole interactions. The best medium for experiments with travelling magnons is thin films of yttrium iron garnet (YIG) epitaxially grown on gadolinium-gallium garnet (GGG) substrates. In these films and for small magnon wave numbers $k$ (0 to 1000 rad/cm), the coupling of ferromagnetic spins is predominantly by a dynamic dipole (or stray) magnetic field they generate collectively while precessing. The total magnon energy is then the sum of energies of the dipole-dipole and Zeeman interactions. The Zeeman interaction results in formation of an energy gap in the magnon spectrum, such that application of a reasonably small external spatially uniform magnetic field $H$ able to saturate the YIG films magnetically shifts the frequency of spin waves $\omega(k=0)$ into the microwave frequency range.

In addition, the dependence of spin wave dispersion $\omega(k)$ on the applied field results in efficient scattering of travelling spin waves from localised non-uniformities of a magnetic field applied to a YIG film [10,11]. A highly localised field non-uniformity can easily be created by placing a narrow metallic wire on top of the film and sending a relatively small dc current through it. The current induces a dc Oersted field around the wire. The Oerstead field is highly localised, thus forming the necessary conditions for efficient spin wave scattering from the Zeemann-energy barrier.

In Ref.[11], it was shown that by controlling the strength of the current, the coefficients of spin-wave transmission and reflection from the localised non-uniformity may be varied in a very broad range – from complete reflection to complete transmission. This potentially creates conditions for forming a semi-transparent mirror for magnons traveling in the film. Importantly, the spin wave dispersion strongly depends on the orientation of the applied field with respect to the film surface and the direction of spin wave propagation in the film plane [12,13]. In order to use the Oerstead field of a wire as a narrow energy barrier, the wire must lie in the film plane and be orientated perpendicular to the SW wave vector, and have a component of its Oersted field parallel to a static spatially uniform magnetic field also applied to the film. These conditions are satisfied for the backward volume magnetostatic spin wave (BVMSW) and the forward volume magnetostatic spin wave (FVMSW). A FVMSW propagates in a film, which is magnetised by applying a static field perpendicular to its plane [13]. A wave has a BVMSW character if a film is magnetised in its plane and the wave propagates along the direction of an applied field [12].

In this work, we focus on creating an electrically controlled mirror for BVMSW. The case of FVMSW is slightly more involved, as these waves scatter from the perpendicular-to-plane component of the Oersted field of a wire. This component of the field has an antisymmetric shape in the direction of the spin-wave wave number. This breaks symmetry of the scattering geometry. Our numerical simulation show that the symmetry break results in slightly different time delays of waves reflected from the barrier if they are incident onto the barrier from its opposite sides. This may affect indistinguishability of single magnons scattered from the mirror.

On the contrary, BVMSW are sensitive to the component of the wire Oersted field that lies in the film plane. This component is symmetric, and hence must preserve indistinguishability of magnons. It is the main reason for the focus on BVMSW in the present work. In the work, we derive an integro-differential equation that governs classical dynamics of envelopes of narrow packets of BVMSW in a ferromagnetic film and simulate excitation, propagation and scattering of SW by numerically solving the envelope equation.

We show that short BVMSW pulses scatter efficiently from a Zeemann-energy barrier created by an Oersted field of a dc current in a wire orientated in the film plane and along the BVMSW wave fronts. By properly adjusting the current through the wire, a semi-transparent SW mirror can be created. The mirror is characterised by a transfer matrix that satisfies requirements for the presence of a Hong-Ou-Mandel Effect for it. We also discuss technical detail how to organise a future experiment on observation of the magnonic Hong-Ou-Mandel Effect based on the proposed semi-transparent SW mirror.

## 2. Theory and simulation results

Fig. 1 shows a sketch of the problem geometry. Its main component is a 5-micron thick ferromagnetic film (film thickness $L$=5 microns). The film is magnetised in its plane by a spatially uniform magnetic field **H** of a strength $\mu_0 H$= 85 mT ($H$=850 Oe). This creates conditions for the existence of BVMSW waves in the film. Two wire-loop transducers (labelled as Port 1 and Port 2), placed $d$=7 mm apart, excite and receive signals carried by an SW. Appendix 3 explains, why we opt for the loop transducers in our simulations and how we simulate the driving microwave magnetic field $h_{dr}$ of a transducer operating as an input one and how we simulate output transducer signals, when a transducer receives a BVMSW signal.

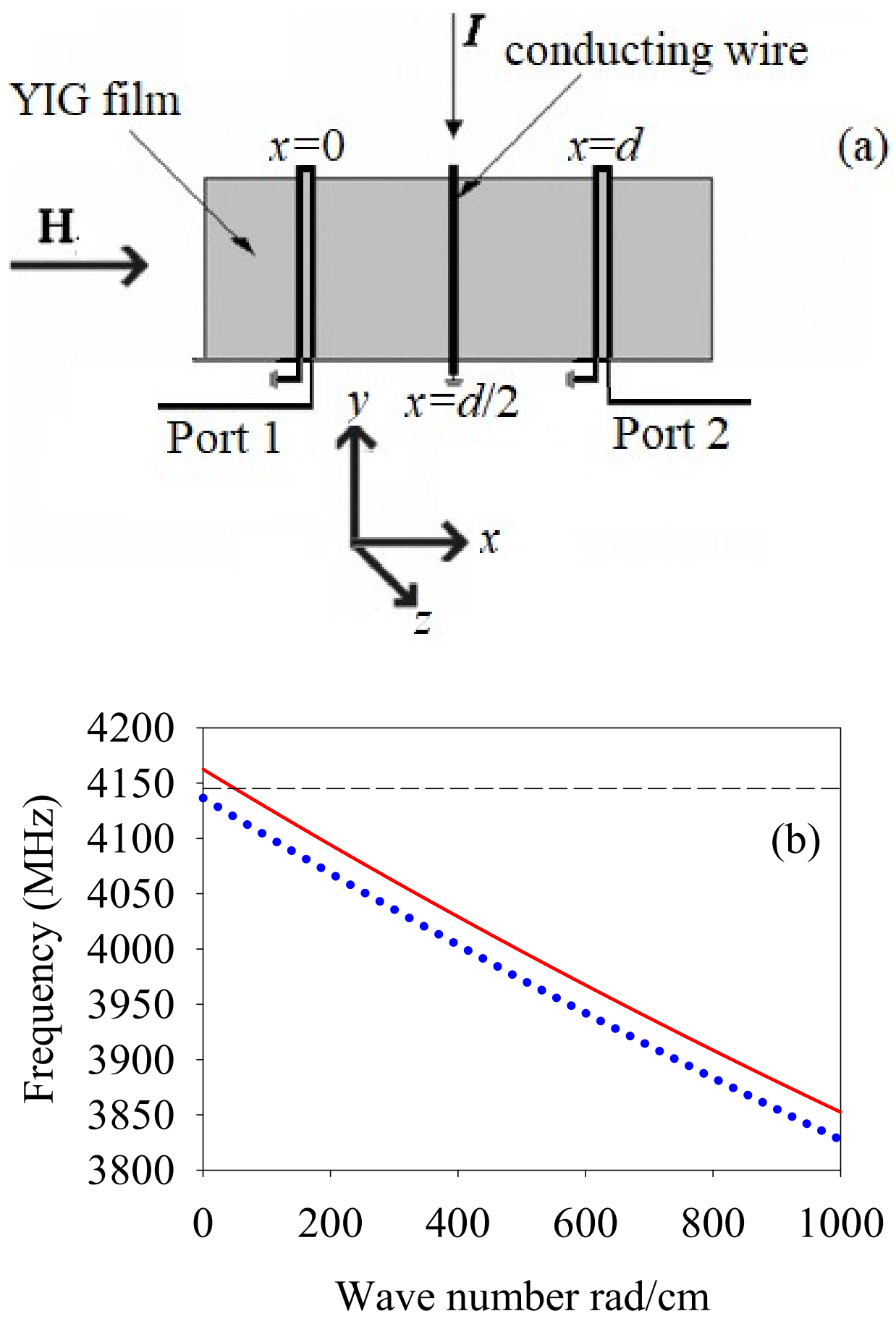

Fig. 1. (a) Sketch of the simulated geometry. (b) BVMSW dispersion for a 5 micron thick YIG film. Solid line: applied field $H$=850 Oe. Dotted line: a slightly smaller applied field $H$=842 Oe. Thin dashed line shows the carrier frequency of the simulated input microwave pulses $\omega_0/2\pi$ = 4145 MHz.

A single wire (shown as "conducting wire" in the sketch) is located midway between the transducers (i.e. at $x$=$d$/2=3.5 mm). The wire width $w$=50 micron, and the wire carries a dc current $I$. The Oersted field of the current $\delta h(I,x)$ creates a potential barrier for SW in the film. A BVMSW pulse is excited by one transducer, or two counter-propagating SW pulses are excited by both transducers simultaneously. The SW pulse(s) travel(s) through the film and scatter(s) from the barrier to create (a) reflected and (a) transmitted pulse(s). The reflected and transmitted pulses are received by the transducers and create output microwave signals of Ports 1 and 2.

As shown in Appendix 1, the BVMSW dynamics in this geometry is governed by an integro-differential equation as follows:

$$dm(x,t)/dt - i\left[\frac{\left(\omega_H + \gamma\delta h(I,x)\right)^2}{2\omega_0} - \frac{\omega_0}{2}\right]m(x,t) + i\frac{\left(\omega_H + \gamma\delta h(x)\right)\omega_M}{2\omega_0}\int_{-\infty}^{\infty}\hat{G}(x-x')m(x',t)dx' = -i\frac{\omega_H\omega_M}{2\omega_0}h_{dr}(x,t) \quad . (1)$$

Here $\gamma$ is the gyromagnetic ratio, which is equal to $2\pi\cdot 28$ MHz/mT ($2\pi\cdot 2.8$ MHz/Oe) for YIG, $\omega_H=\gamma(H+i\Delta H)$, $2\Delta H=0.5$ Oe is the magnetic loss parameter for YIG (at room temperature), $i$ is imaginary unit, $\omega_M=\gamma M$, where $M$ is films' saturation magnetisation ($\mu_0 M=175$ mT for YIG), and $m(x,t)$ and $h_{dr}(x,t)$ are the perpendicular-to-plane vector components of the slow envelopes of dynamic magnetisation of BVMSW and of the microwave driving field respectively. They are slow with respect to the microwave carrier frequency $\omega_0$ of these waveforms. The expressions for the Green's function of the perpendicular-to-plane component of the dipole magnetic field of the dynamic magnetisation $\hat{G}(s)$ [14] and for the in-plane component of the Oersted field $\delta h(I,x)$ of the wire that creates the potential barrier are given in Appendices 1 and 2 respectively.

In this work, we solve the envelope equation numerically. We use a mesh of 1000 discrete co-ordinate points $x=x_n$, $n = 1,2,..1000$. This creates a 1000-component vector with components $m(x_n,t)$. We employ the fourth-order Runge Kutta method to obtain a time dependent solution for the vector. The whole solution process is implemented as a MathCAD® worksheet employing MathCAD's Adaptive Runge-Kutta solution function *Rkadapt*. Simulating the whole process of pulse scattering from the barrier takes 3 to 5 minutes of computer time.

Similar to the experiment on the phononic HOME effect [4], we assume that microwave superconducting qubits act as sources of single microwave photons, which the transducers convert into single magnons. The photons were formed as 17.6 ns long pulses in [4]. Similarly, we use 20 ns long rectangular-shape pulses as $h_{dr}(t)$.

All simulations in this work are carried out for a carrier wave number of SW of 49 rad/cm. This corresponds to $\omega_0$=4145 MHz, as shown in Fig. 1(b). Fig. 2 displays the simulated "screenshots" of a BVMSW pulse launched by the transducer located at $x$=0 (Port 1). The screenshots "are taken" at different moments of time $t$. No potential barrier is present ($I$=0). $t$=0 corresponds to the front edge of the driving pulse $h_{dr}(t)$. The transducer excites two BVMSW pulses travelling in two opposite directions from it, but we only show the pulse propagating towards the receiving antenna (located at $x$=7 mm).

The waveform shape evolves during the pulse propagation in the film. The shape of the envelope of the BVMSW pulse is not rectangular even for small $t$. This is because the pulse edges are smoothed during the transduction process due to the finite width of the transduction frequency/wave-number band of the transducer. As the pulse propagates in the film, it broadens and its height drops. The former is an effect of the presence of dispersion, which distinguishes spin wave pulses from surface acoustic waves exploited in [4]. The dispersion law for the fundamental mode of BVMSW is shown in Fig. 1(b). One sees that the eigen-frequency vs. wave number dependence for the waves does not represent a straight line – the line has a slight positive curvature. This makes narrow pulses carried by BVMSW prone to dispersion broadening [15].

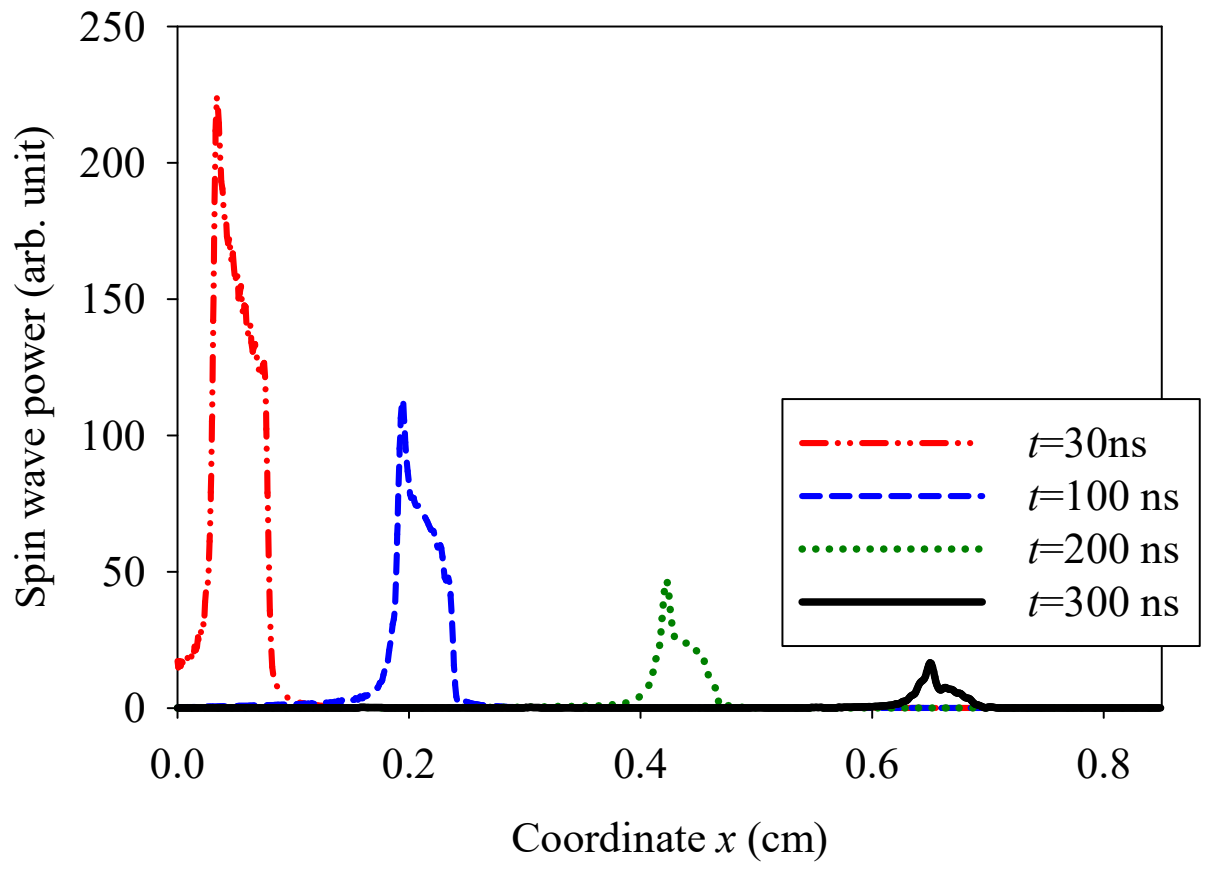

Fig. 2. Snapshots of BVMSW pulses taken at different moments of time $t$ (see the legend). Carrier wave number of pulses is 49 rad/cm, which corresponds to $\omega_0/2\pi$ = 4145 MHz. Applied field $H$=850 Oe. Film thickness $L$=5 micron. Length of input microwave pulses: 20 ns. The pulses are excited by the transducer located at $x$=0. No potential barrier is present ($I$=0).

The decrease in the pulse amplitude has two origins. The first one is the presence of magnetic loss in the medium. Due to the loss, the pulse amplitude drops exponentially with distance from the location of pulse excitation. The second origin is the conservation of pulse energy - when a pulse broadens in a dispersive medium, its height must decrease, in order to conserve energy of the pulse.

Figure 3(a) demonstrates exemplary snapshots of the same BVMSW pulse taken in the presence of a potential barrier. $I$=–0.35 A for the example. The negative sign indicates that within the film, the Oersted field of the current is anti-aligned to the applied field **H**. As a result, the total static field within the barrier is smaller than elsewhere. This regime of SW transmission through a barrier was termed "tunnelling" in Refs. [10,11]. The decrease in the total static field locally shifts the dispersion relation downwards in frequency. If the shift is small, this just makes the local value of the BVMSW wave number smaller. However, for larger magnitudes of the negative $I$, the downshift of the dispersion relation becomes large enough to place the BVMSW frequency locally into the energy gap that exists *above* the BVMSW band. This situation is shown in Fig. 1(b) by the dotted line. One sees that for a smaller $H$, any point of the dispersion relation lies below $\omega_0$ (shown by the horizontal dashed line). Then the only way for a BVMSW to cross the barrier is by tunnelling through it as a leaky wave.

The example from Fig.3(a) is for a specific value of $I$, for which energy of the transmitted (the right-hand-side waveform) pulse equals to energy of the reflected pulse (the left-hand-side pulse). The quantity we actually plot in the graph is $|m(x,t)|^2$. Power carried by a spin wave must scale as $|m(x,t)|^2$. Therefore, in the following, we term this quantity "SW power". Similarly, integrating $|m(x,t)|^2$ over the spatial or temporal width of the pulse we obtain a quantity, which scales as pulse energy. In the following, we will term the integral "pulse energy".

In Fig. 3(a) the barrier is located at $x$=0.35 cm. One sees that the scattered pulses are located at equal distances from this point. This demonstrates that the pulses acquire the same time delay while scattering from the barrier. This is better seen from Fig. 3(b), which displays microwave signals received by the transducers. The signal from Port 1 is the reflected signal, and the one exiting Port 2 is the transmitted one. One sees that the shapes of the output pulses are slightly different, but they overlap quite well in time. The same power vs. $t$ traces as in Fig. 3(b) are obtained when we excite an SW pulse with the second transducer (Port 2) located at $x$=0.7 cm (not shown). The only difference is that the reflected pulse is now received by Port 2 and the transmitted one by Port 1.

Panel (c) of the same figure displays the output pulses of the ports for a case, when two identic rectangular microwave pulses are applied to both ports simultaneously. The only difference between the input pulses is their initial phases. The phase difference between the input pulses is either $+\pi/2$ or $-\pi/2$. We see that in either case, the microwave signal from one port is strong, and the output signal of the other port is strongly suppressed. This behaviour is in full agreement with what is expected from an optical semi-transparent mirror [2]. We also see that changing the phase difference by $\pi$ (i.e. from $+\pi/2$ to $-\pi/2$) swaps ports, through which the larger and the smaller pulses exit. Again, this is in agreement with what is expected from an optical semi-transparent mirror [16].

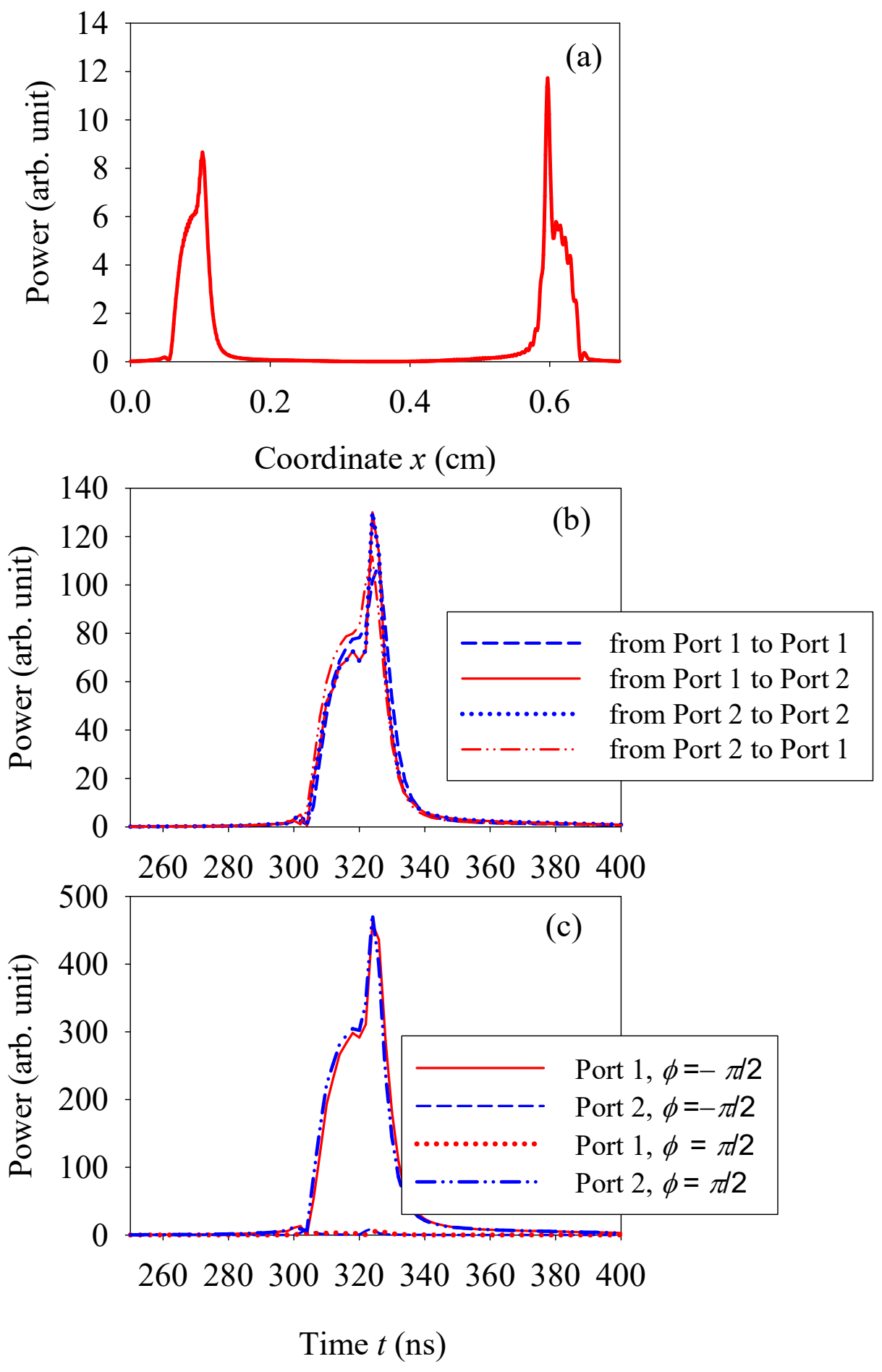

Fig. 3. (a) Snapshot of BVMSW pulses scattered from a barrier $I = -0.35$ A. The snapshot corresponds to $t$ =140 ns. (b) Output microwave signals of Port 1 (denoted as “to Port 1” in the legend) and Port 2 (denoted as “to Port 2”), when a single input microwave pulse is applied to either Port 1 (denoted as “from Port 1”) or Port 2 (denoted as “from Port 2”). The barrier height is the same as for Panel (a). (c) The same as (b) but the same microwave pulses are now applied to both ports simultaneously. “Port 1” and “Port 2” in the legend denote the ports that receive the respective signals. $\phi$ is the initial phase of the input microwave pulse applied to Port 2. The initial phase of the input pulse fed into Port 1 is zero. The barrier height is the same ($I = -0.35$ A). The other simulation parameters are the same as for Fig. 2.

Figure 3(c ) is the main finding of this work. It demonstrates that the barrier acts as semi-transparent mirror, for which we may expect a magnonic HOME effect. But before we proceed to a discussion of the effect, it is worth noting that the mirror transparency is electrically controlled. The ratio *R* of transmitted and reflected pulse energies can easily be decreased or increased by altering *I*. This is shown in Fig. 5(a) in Appendix 4.

## 4. Magnonic HOME

Important for a theoretical description of HOME is establishing the transfer matrix for a semi-transparent mirror. The general form of the matrix is

$$\hat{S} = \begin{pmatrix} \dot{S}_{11} & \dot{S}_{12} \\ \dot{S}_{21} & \dot{S}_{22} \end{pmatrix}. \quad (2)$$

The matrix element $\dot{S}_{11}$ represents the signal exiting Port 1 if an input pulse of amplitude of 1 is applied to Port 1. Similarly, $\dot{S}_{22}$ is the output signal of Port 2, if an input pulse is applied to Port 2. Physically, these quantities represent reflection coefficients from the barrier. $\dot{S}_{21}$ and $\dot{S}_{12}$ represent the coefficients of transmission of the mirror – from Port 1 to Port 2 and from Port 2 to Port 1 respectively. The dots on top of the symbols *S* emphasise that we deal with complex-valued quantities. BVMSW is a reciprocal wave (has the same properties while propagating in +*x* and –*x* directions); therefore, we expect that $\dot{S}_{22} = \dot{S}_{11} = \dot{\tau}$ and $\dot{S}_{12} = \dot{S}_{21} = \dot{r}$, where we introduced the complex-valued transmission $\dot{\tau}$ and reflection $\dot{r}$ coefficients. Our numerical simulations results are in excellent agreement with these equalities.

We will establish the matrix specifically for the case of a perfectly semi-transparent mirror and for the pulse regime we studied above. This implies that energy of the transmitter pulse equals energy of the reflected one ($|\dot{\tau}|^2 = |\dot{r}|^2$). This is the case of *I*=–0.35 A. We will also neglect losses of the pulses due to magnetic damping on their way to and from the barrier, thus, effectively moving the ports to the "edges" of the barrier. In addition, we will neglect the signal loss within the barrier due to the same magnetic damping. As the barrier is short, this loss is insignificant. Under these conditions, and taking into account the energy conservation law ($|\dot{\tau}|^2 + |\dot{r}|^2 = 1$) we obtain $|\dot{S}_{11}| = |\dot{S}_{21}| = |\dot{S}_{22}| = |\dot{S}_{12}| = |\dot{\tau}| = |\dot{r}| = \frac{1}{\sqrt{2}}$.

We also need phases of these quantities. To find the phases, in Fig. 4 we plot the phase differences $\Delta\phi_1 = \arg(\dot{S}_{21}) - \arg(\dot{S}_{11})$ and $\Delta\phi_2 = \arg(\dot{S}_{12}) - \arg(\dot{S}_{22})$, where arg(*z*) denotes the phase of a complex number *z*. We obtain the phase differences from simulating single pulses incident onto the barrier either from Port 1 or 2. The data are shown for the case of the perfectly semi-transparent BVMSW mirror (*I*=–0.35 A). This is the configuration of Fig. 3(b). Calculating the differences eliminates phase accumulation by the signals on their paths to and from the barrier. Again, this approach is consistent with moving the ports to the edges of the barrier.

For convenience, we also show the simulated $|\dot{S}_{21}|^2$ in the same graph. One sees that both phase differences are very close to $\pi/2$ within the pulses. This explains why phase differences of $\pm\pi/2$ between the input pulses switch the direction of the large-amplitude output pulse (Fig. 3(c)). By adding or subtracting $\pi/2$ to either $\Delta\phi_1$ or $\Delta\phi_2$, we create conditions for constructive interference of scattered pulses incident onto one port and for destructive one for pulses travelling towards the other port.

Changing the sign of this extra phase shift swaps the interference conditions for the ports and sends the large-amplitude pulse in the opposite direction of the axis $x$ with respect to the original one.

We may assign the phase $\Delta\phi_1=\Delta\phi_2=\pi/2$ either to the diagonal ( $\dot{r}$ ) or anti-diagonal ( $\dot{\tau}$ ) elements of the matrix, as we are not interested in the absolute phase accumulation by the signal while being scattered from the barrier, but just in the phase difference between the transmitted and reflected wave, given the reciprocal character of the mirror. Assigning the phases to the transmission coefficients yields a transfer matrix as follows

$$\hat{S} = \frac{1}{\sqrt{2}}\begin{pmatrix} 1 & i \\ i & 1 \end{pmatrix}. \quad (3)$$

The matrix has the same form as the transfer matrix for the "symmetric beam splitter of Loudon" [3,16]. This finding implies that the formalism from Refs. [2,4] applies to our spin wave mirror. It allows calculating the output quantum state of the mirror for the case when single magnons are incident on Port 1 and Port 2. If two bosons arrive simultaneously to two detectors, they are called "coincidence bosons". Consider coincidence magnons. These magnons must be fully indistinguishable. From Fig. 3(b) we clearly see that the output pulses of the mirror overlap in time quite well. Hence, coincident arrival of two single magnons, shaped as those pulses, to the two ports is possible, provided they have been excited simultaneously. Hence, in figures 3(b) and 3(c) we deal with coincidence magnons.

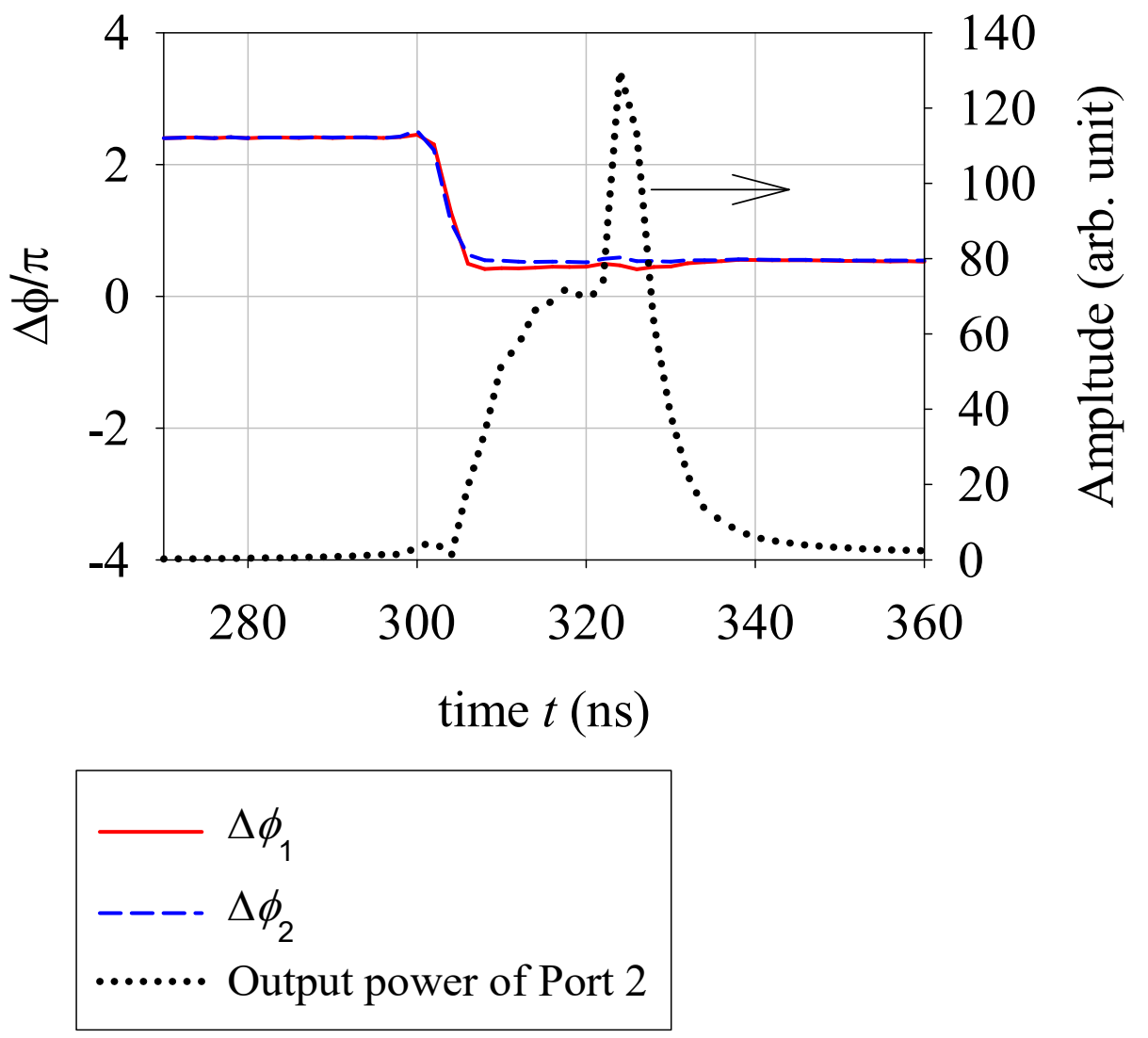

Fig. 4. Thick solid line: difference in phases $\Delta\phi_1$ of signals received by Port 1 and Port 2, when a single input pulse is applied to Port 1. Dashed line: the same, but the single input pulse is now applied to Port 2 ($\Delta\phi_2$). Dotted line: microwave output power of Port 2, when a single input pulse is applied to Port 1 (right-hand vertical axis). $I$=–0.35 A. The other simulation parameters are the same as for Fig. 2.

We assume that two single microwave photons are incident simultaneously on Ports 1 and Port 2 to create to coincident initial magnons. We will denote this state $|1,1>_{in}$. The reflected and transmitted pulses from Fig. 3(b) have very similar shapes. Therefore, for simplicity, we may assume that the respective single-photon-state spectral amplitude functions [2,4] are identical. In addition, the pulses from Fig. 3(b) overlap very well in time. Therefore, it is reasonable to assume that the relative time delay of arrival of the pulses to the transducers [4] vanishes. Under these assumptions, the last equation from Section A of online supplemental materials for Ref.[4] reduces to simple formulas from Table 1 in [3].

As follows from the third cell of the third column of the table in Ref.[3], the probability $P_{11}$ of the observation of single microwave photons simultaneously exiting both ports (the output state |1,1>$_{out}$) is given by

$$P_{11} = |(\dot{\tau}|^2 - |\dot{r}|^2)^2 = (1-2|\dot{r}|^2)^2 = \left[1-R^2\right]^2 / \left[1+R^2\right]^2, \quad (3)$$

where $R = |\dot{\tau}/\dot{r}|$. One sees that for a perfectly semi-transparent mirror ($|\dot{\tau}|^2 = |\dot{r}|^2 = 1/2$), this formula reduces to $P_{11}$=0. This implies that the output state of the system is either |0,2>$_{out}$ or |2,0>$_{out}$. Each of these states is characterised by the presence of both coincidence photons at the same port – either Port 2 or Port 1 respectively. The probabilities $P_{02}$ and $P_{20}$ to observe these states are given by the second and the fourth cells of the same column of the table. Both probabilities are the same and are equal to $2|\dot{\tau}|^2|\dot{r}|^2$, which reduces to $P_{02}$ = $P_{20}$=1/2 for the semi-transparent mirror.

Note that Eq.(3) is valid for a lossless medium. Because of the very short length of the potential barrier, it is acceptable to neglect magnetic losses in the mirror. However, we cannot neglect propagation losses of SW on its path from an input port to an output one. Our simulation for *I*=0 shows that for *d*=7 mm, the BVMSW propagation losses are 18 dB, which makes 2.6 dB/mm. This signal attenuation does not include transducer impedance mismatch loss (See e.g. [17]). In physical experiments, transducers are not well impedance matched usually. Therefore, we may expect 5 to 10 decibel of loss of conversion of power of an input microwave signal into a SW signal by an input transducer. The same then applies to the back conversion by an output transducer. In addition, the input transducer will lose extra 3 dB of energy for exciting a BVMSW pulse, which travels in the –*x* direction. This loss is usually termed "bi-directionality loss". The bi-directionality loss also takes place at the output transducer - only half of SW power incident onto the output transducer contributes to generating the output microwave signal. Hence, the total insertion loss of our device is 18 dB + 2x10 dB + 2x3 dB = 44 dB, if we use the conservative estimation of the transducer conversion losses of 10 dB per transducer. Note that we did not have to choose *d*=7 mm. Actually, a smaller transducer separation of *d* =4 mm usually works fine in SW experiments and technology. For *d*=4mm, we will have (2.6x4) dB + 2x10 dB + 2x3d B $\approx$ 36 dB. This implies that for *d*=4 mm and *I*=0, we will have $|\dot{\tau}|^2 = 10^{-36/10} = 2.5\cdot 10^{-4}$. Then, from the energy conservation law, for a semi-transparent mirror (*I*=–0.35 A) we must have $|\dot{\tau}|^2 = |\dot{r}|^2 = 2.5\cdot 10^{-4}/2 = 1.25\cdot 10^{-4}$.

In Ref. [18] it was shown that the formalism from Ref.[3] remains valid in the presence of losses in a medium, provided that $\dot{\tau}$ and $\dot{r}$ remain orthogonal ($\dot{\tau} = i\dot{r}$) in the presence of a loss, and we count coincidence output states only. As seen from Fig. 4, the condition $\dot{\tau} = i\dot{r}$ is satisfied for the semi-transparent BVMSW mirror ($|\dot{\tau}|^2 = |\dot{r}|^2$). The theory from Ref. [18] shows that for the input state |1,1>$_{in}$ and $|\dot{\tau}|^2 = |\dot{r}|^2$, two boson detectors never click simultaneously – either one of the detectors clicks and the other remains silent ("non-co-incident" photon states |1,0>$_{out}$ and |0,1>$_{out}$,) or one detector displays a signal of simultaneous arrival of two bosons, but the other detector remains silent (states |0,2>$_{out}$ or |2,0>$_{out}$).

Equations (3.12) from [18] express probabilities of observing the output states of a lossy mirror. Specifically for the perfectly semi-transparent lossy magnon mirror (*I*= –0.35 A), $\dot{\tau} = i\dot{r}$ and complete temporal overlap of output pulses (Fig. 3(b)), the probability of one quasiparticle to survive $P_{10}$=$P_{01}$≈ $2|\dot{r}|^2 = 2\cdot 1.25\cdot 10^{-4} = 2.5\cdot 10^{-4}$. The probability of observing no microwave photon at either port (the output state |0,0>$_{out}$) $P_{00} = \left(1-2|\dot{r}|^2\right)^2 \approx 1$. Similarly, the probability to observe one microwave photon

at each port $P_{11}$= 0, and the probabilities for both photons to exit the same port are $P_{02}$ =$P_{20}$= $2|\dot{r}|^4 = 3.1\cdot 10^{-8}$.

Assume that we are able to apply and detect single microwave photons to the device with a repetition period of 10 microseconds. Then we may expect one detection of either state |0,2>$_{out}$ or |2,0>$_{out}$ state every 30 seconds or so. This will allow collecting reliable statistics within a couple of hours of running the experiment.

Note that we need count coincidence photons only (we will call this "coincidence events" in the following) and ignore all |1,0>$_{out}$ and |0,1>$_{out}$ states. However, while acting as microwave photon detectors, some qubits may be able to absorb just one microwave photon and unable to absorb two photons [4]. Therefore, the experimentalists had to rely on an indirect method of observing the |0,2>$_{out}$ and |2,0>$_{out}$ states. They counted the |1,1>$_{out}$ states as a function of the degree of indistinguishability of the single photons and saw that this dependence has a sharp minimum for the highest degree of the phonon indistinguishability. We believe that the same method can also be used for the BVMSW magnons. Furthermore, one can use the same way of controlling magnon distiguishibility – by delaying the launch of one of the single magnons into the film. One expects to see a minimum in the number of |1,1>$_{out}$ coincidence events for vanishing time delay. The observation of the drop in $P_{11}$ will evidence that |2,0>$_{out}$ and |0,2>$_{out}$ states are created instead of the |1,1>$_{out}$ states.

Alternatively, one may exploit the electric control of transparency of the SW mirror. Set $I$ to 0 first and measure the $P_{11}$ and $P_{10}$+$P_{01}$. Then turn on $I$, set the mirror transparency to 50% by properly adjusting the current and repeat the experiment. Observation of a decrease in $P_{11}/(P_{10}+P_{01})$ will then evidence the presence of the magnonic HOME effect. (Because magnetic damping in the barrier may be different from elsewhere, it may be worth normalising $P_{11}$ by $P_{10}$+$P_{01}$, as we did in the formula above, in order to account for the small potential change in the overall level of magnetic losses in the presence of the barrier.)

It may also be worth measuring $P_{11}$ as a function of $I$. Panel (b) of Fig. 5 from Appendix 4 shows the $P_{11}(I)$ dependence calculated with Eq.(3) using the values of $R(I)$ from Fig. Fig.5 (a) and assuming that the phase difference between $\dot{\tau}$ and $\dot{r}$ is $\pi/2$ for any $I$. We checked the latter assumption with numerical simulations for a range of $I$ values from −1 A to −0.1 A. We found that it is satisfied with a good accuracy over the whole range. The graph in Fig. 5(b) is characterised by three zeros of $P_{11}(I)$ corresponding to three $I$ values, for which $R(I)$ =1 (see Fig. 5(a)). Observation of the zeros (or perhaps just minima in a real-life experiment) will also evidence the presence of a magnonic HOME.

## 5. Conclusion

Through numerical simulations employing an original formalism we demonstrated that a potential barrier for magnons, created by an Orsted field of a dc current flowing through a wire sitting on top of a YIG film, is able to act as an electrically controlled partly transparent mirror for the magnons. We considered the backward volume magnetostatic wave (BVMSW) configuration specifically and found that the mirror transparency can be set to 50% ($R$=0.5) by properly adjusting the current strength, thus creating a semi-transparent mirror. The strongest Hong-Ou-Mandel Effect is expected for single magnons in the $R$=0.5 configuration. The effect must be seen as both coincidence magnons, launched into the film from two transducers (ports) located from the opposite sides of the mirror, simultaneously arriving to the same transducer and creating a two-microwave-photon state at the transducer output port. The probabilities of seeing those two-photon states at either output port must be the same. This result applies to coincidence magnons only. Due to significant transmission loss of the device, most of detection events will be of detecting no output photons.

Appendix 1. Derivation of Eq.(1)

Following the idea from [11], we may write

$$\chi^{-1}(\omega)\,\tilde{m}(x,\omega)=\int_{-\infty}^{\infty}\hat{G}(x-x')\tilde{m}(x',\omega)dx'+\delta h(I,x)+\tilde{h}_{dr}(x,\omega), \quad (4)$$

where

$$\chi(\omega)=\omega_H\omega_M/(\omega_H^2-\omega^2), \quad (5)$$

and $\chi^{-1}(\omega)=(\omega_H^2-\omega^2)/(\omega_H\omega_M)$ accordingly, $\omega_H=\gamma(H+i\Delta H)$, $H$ is a spatially homogeneous applied field, $i$ is imaginary unit, $\omega_M=\gamma M$, $M$ is films' saturation magnetisation, $\gamma$ is gyromagnetic ratio, $\delta h(I,x)$ is a spatially localised Oersted field of a wire that acts as a potential barrier for spin waves, $I$ is a dc current through the wire that creates the Oersted field, and $\tilde{m}(x,\omega)$ and $\tilde{h}_{dr}(x,\omega)$ are the perpendicular-to-film-plane vector components of the dynamic magnetisation of BVMSW and of the microwave driving field respectively. Both have been averaged over the film thickness $L$ and are assumed to be oscillating at a microwave frequency $\omega$. In order to account for magnetic losses in the YIG film, we add an imaginary part $i\Delta H$ to the applied field $H$ while calculating the parameter $\omega_H$ [19]. Here $\Delta H$ is the material's magnetic loss parameter.

$\hat{G}$ is the Green's function of the perpendicular to plane dipole field of dynamic magnetisation [14] averaged over the film thickness $L$

$$\hat{G}(s)=\frac{2}{L}\ln\left[\frac{s^2}{s^2+L^2}\right], \quad (6)$$

where $s$ is a dummy variable.

We now perform an inverse Fourier transform of both sides of Eq.(4) with respect to $\omega$. We obtain

$$\int_{-\infty}^{\infty}\exp(i\omega t)\tilde{m}(x,\omega)d\omega=\tilde{m}(x,t), \quad \int_{-\infty}^{\infty}\exp(i\omega t)\tilde{h}_{dr}(x,\omega)d\omega=\tilde{h}_{dr}(x,t)$$

and

$$-\int_{-\infty}^{\infty}\omega^2\exp(i\omega t)\tilde{m}(x,\omega)d\omega=\partial^2\tilde{m}(x,t)/\partial t^2,$$

where $t$ is time. We then assume $\tilde{h}_{dr}(x,t)=h_{dr}(x,t)\exp(i\omega_0 t)$, where $h_{dr}(x,t)$ is a slow envelope of the waveform of the driving field $\tilde{h}_{dr}(x,\omega)$, and $\omega_0$ is its microwave carrier frequency. Given the assumed form of the driving term, we assume a similar solution for the dynamic magnetisation: $\tilde{m}(x,t)=m(x,t)\exp(i\omega_o t)$, where $m(x,t)$ is a slow envelope of the waveform of $\tilde{m}(x,\omega)$. Upon substituting this solution into the Fourier transformed Eq.(4) and ignoring an emerging $\partial^2 m(x,t)/\partial t^2$ term as negligibly small with respect to $\omega_0^2 m(x,t)$, which also appears in the equation, we obtain Eq.(1) from the main text of the paper. In addition, while writing down Eq.(1) we take into account that $h_{dr}(x,t)$ and $\delta h(I,x)$ are highly localised and do not overlap spatially.

Appendix 2. Equation for $\delta h(I,x)$

We assume that the barrier $\delta h(I,x)$ is created by the Oersted field of a wire with a circular cross-section of a radius $r$. The wire lies directly on the surface of the YIG film. We are interested in the component of the Oersted field that is in the plane of the film. Employing Ampere circuital law and averaging the so-obtained in-plane component of the Oersted field over the film thickness we obtain

$$\delta h(I,x) = \frac{1}{4\pi L} \ln\left[\frac{x^2+(r+L)^2}{x^2+r^2}\right]. \quad (7)$$

Appendix 3. Simulating magnon excitation and detection by the wire-loop transducers

BVMSW waves couple to the perpendicular-to-plane component of the microwave driving field. From the computational point of view, it is more convenient to assume wire-loop transducers in the form of one period of a meander line. This creates a current loop. (In a real-world experiment, simpler single-microstrip transducers will potentially be more convenient.) The advantage of the wire-loop transducers for the simulations is that the perpendicular component of their Oersted field is symmetric with respect to transducer's symmetry axis and strongly localised within the current loop. The focus of the paper is not on BVMSW excitation and detection. Therefore, for simplicity, we may assume that the driving field $h_{dr}$ is uniform within the current loop (i.e. from $x= 0 - w_a/2$ to $0 + w_a/2$, where $w_a$ is the width of the loop in the direction $x$) and vanishes elsewhere. The same applies to the second transducer located at $x$=7 mm.

The same transducers may be used to receive the BVMSW signals. The operation of a receiving transducer is based on Faraday induction (see e.g. [17]). Therefore, the output microwave voltage of the transducer must scale as microwave magnetic flux through the loop of the transducer. A good proxy for the latter is the BVMSW amplitude $m(x,t)$ integrated over the width of the loop $w_a$. The output power of the transducer then scales and square of the flux and hence as modulus square of integral of $m(x,t)$ over the transducer loop width. We use the latter quantity as a proxy to output microwave power of the transducers. Energy of the output pulse is then obtained by integrating the output microwave power over the temporal width of the output pulse.

Appendix 4: Dependence of the ratio of transmitted energy to reflected energy

The dependence of the ratio of transmitted energy to reflected energy $R(I)$ is obtained by solving Eq.(1) numerically, specifically for 20-nm long microwave input pulses. It is displayed in Fig. 5(a). The graph shows $R$ as ratio of energies of pulses received by the respective ports. The dashed line in the figure sows the level of $R$=1. Figure 5(b) displays the respective probability $P_{11}$ to observe coincident single photons exiting Port 1 and Port 2 simultaneously (state $|1,1>_{out}$) as a function of $I$, when the input state of the device is $|1,1>_{in}$. The three points, where $P_{11}$ vanishes, correspond to $R$=1. To produce this graph, we used Eq.(3) from the main text of the paper.

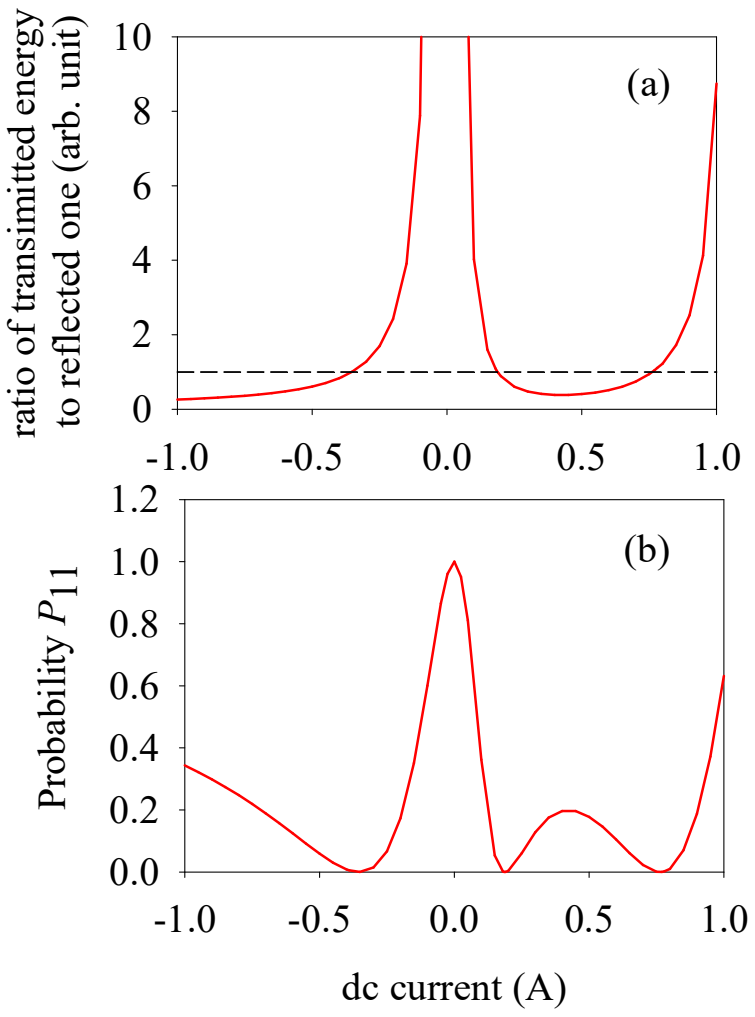

Fig. 5. (a) Ratio $R$ of energy of the pulse received by Port 2 to energy of the pulse received by Port 1 as a function of the current $I$ creating the potential barrier. The input microwave pulse is applied to Port 1. (b) Probability to observe the quantum state $|1,1>_{out}$ if the input quantum state of the device is $|1,1>_{in}$ as a function of $I$.